\documentclass{article}
\usepackage{spconf,amsmath,graphicx}
\usepackage{relsize}
\usepackage[table,xcdraw]{xcolor}
\usepackage{soul}
\usepackage{url}
\usepackage{subfigure}
\usepackage{mathtools}
\usepackage{booktabs}
\usepackage{enumitem}
\usepackage{amssymb}
\usepackage{tabularx}

\title{Efficient Keyword Spotting by capturing long-range \\interactions with Temporal Lambda Networks}
\name{Biel Tura$^{1,2}$, Santiago Escuder$^{1,2}$, Ferran Diego$^2$, Carlos Segura$^2$, Jordi Luque$^2$}
\address{   $^1$Universitat Politècnica de Catalunya \\
            $^2$Telefónica I+D, Research, Spain}
\begin{document}
%
\maketitle
\begin{abstract}
Models based on attention mechanisms have shown unprecedented speech recognition performance. However, they are computationally expensive and unnecessarily complex for keyword spotting, a task targeted to small-footprint devices.
This work explores the application of Lambda networks, an alternative framework for capturing long-range interactions without attention, for the keyword spotting task. We propose a novel \textit{ResNet}-based model by swapping the residual blocks by temporal Lambda layers. Furthermore, the proposed architecture is built upon uni-dimensional temporal convolutions that further reduce its complexity.
The presented model does not only reach state-of-the-art accuracies on the Google Speech Commands dataset, but it is 85\% and 65\% lighter than its Transformer-based (KWT) and convolutional (Res15) counterparts while being up to $100\times$ faster. To the best of our knowledge, this is the first attempt to explore the Lambda framework within the speech domain and therefore, we unravel further research of new interfaces based on this architecture. 
\end{abstract}
\vspace{-0.15cm}
\section{Introduction}
\vspace{-0.10cm}

Speech recognition is focused on the translation of human speech into understandable text. This task is usually tackled through Convolutional Neural Networks (CNN) and Sequence-to-Sequence models (Seq2Seq). The former are efficient approaches for both extracting local-time dependencies in speech and for low latency streaming applications \cite{speech_recognition_cnn}. The latter, are inspired by language translation tasks mainly comprising Transformer blocks \cite{attention}. They are based on attention mechanisms that easily capture long-time dependencies of acoustic sequences from larger contexts \cite{wav2vec2, speechBERT, end2end_asr, transformers_asr} or even combining both approaches in hybrid models (CNN-Seq2Seq) known as Conformers \cite{conformer}. Nonetheless, these models are computational and memory expensive and, although they can be adapted to perform the keyword spotting task (KWS), they are far from the optimal solution for deployment under a very limited resources scenario. In practical applications, KWS is preferably deployed as a stand-alone application thus avoiding the uploading of private data to the cloud or edge services from the smart home, IoT or embedded/mobile devices. Due to that fact, any proposed solution for such devices, where the computation capability becomes scarce, shall be accurate enough for the detection of a predefined list of keywords while being small in its memory footprint and computationally cheaper as possible, e.g. measured in terms of parameters and operations respectively. Indeed, this trade-off between accurate detection and low computational resources is an active research field within the speech recognition domain \cite{dcase2021}. Latter advances reported the residual temporal convolution networks as an effective candidate for small-footprint KWS and have been proposed in \cite{cnn_kws, ResNet1d, aura} to address this task.

In this work, we propose to adopt the Lambda layer architecture by Bello \cite{lambda_networks} for the keyword spotting task. Motivated by recent results in speech recognition models with the transformer architecture \cite{speechBERT, kwt}, we propose to capture long-range interactions for the keyword detection problem. Models adopting the transformer architecture for this purpose are computationally complex, however, Lambda layers present a general framework for capturing long-range interactions efficiently, bypassing the need of computing attention maps, through linear functions known as lambdas. Moreover, we propose this framework with temporal uni-dimensional signals, thus drastically reducing the number of operations in the network. This approach was early explored by Cerisara et al. \cite{multiband} by separating different Mel spectrogram bands for a phonetic speech recognition approach and was then used by a few authors for the keyword spotting task \cite{ResNet1d, temporal_conv_kws}. 

Inspired by the current state-of-the-art architectures for keyword spotting based on residual convolutions \cite{res15} and the proposed implementation of the Lambda operator in \cite{lambda_networks}, we introduce a novel architecture for the acoustic keyword spotting based on combining previous two approaches: \textit{LambdaResNet}, an architecture able to reach close to state-of-the-art figures for the keyword detection task without compromising both the model size and the computational complexity. Moreover, the proposed network is even smaller, in terms of parameters and floating point operations, than the smallest proposed models in literature as \cite{ResNet1d}, based on uni-dimensional convolutions.

\section{Related work}
Keyword spotting through neural networks had been tackled with several methodologies. Chen et al. \cite{mlp_kws} and Wang et al. \cite{mlp_kws_temp} built different models parameterized by a multi-layer perceptron architecture that surpassed previous approaches based on Hidden Markov models \cite{hmm}. Sainath and Parada \cite{cnn_kws} introduced the convolutional neural network for the keyword detection problem, outperforming previous works and reducing the network footprint size. 

The usage of recurrent neural networks for the keyword spotting task has also been studied in the work by Arik et al. \cite{rnn_kws}. The added value of recurrent networks is the ability to learn long-range dependencies. However, they are difficult to integrate into real-time speech recognition systems which eventually is the end goal of this task.

More recently and due to the change of paradigm introduced by the self-attention approach in the natural language processing domain \cite{attention}, attention mechanisms had also been studied in the keyword spotting task \cite{att_kws, mhatt_rnn_kws}. These architectures, while matching state-of-the-art performance, they have a larger number of parameters than traditional two-dimensional residual convolutional networks.

Even though models based on attention mechanisms are promising, the keyword spotting problem has moved towards using residual convolutional neural networks mainly to achieve lighter and faster models for the end task. At the moment, the best architectures for the keyword spotting task are based on two-dimensional convolutional networks by Tang et al. \cite{res15}. In addition to this, Vygon et al. \cite{res15_triplet} showed a training approach based on a triplet loss function and phonetic similarity that improves the classification accuracy for the keyword detection using the same architecture. On the other hand, lighter models based on uni-dimensional convolutions following the residual architecture have also been studied by Choi et al. \cite{ResNet1d}, achieving similar metrics than two-dimensional convolutional networks while decreasing the footprint size of the model.

Very recently, a new architecture based on the Vision Transformer (\textit{ViT}) \cite{ViT} has surpassed previous approaches for the Keyword Spotting task. This new model, proposed by Berg et al. \cite{kwt}, is known as the Keyword Transformer (\textit{KWT}) and it is fully based on a self-attentional architecture. The Keyword Transformer outperforms previous works at the cost of very large and computational complex models, not desirable for the end goal of this task in which models are required to run in low-resource devices.

As of our knowledge, we are the first to implement long-range interactions through the recently presented Lambda architecture in the speech domain as well as introducing the uni-dimensional lambda architecture for a specific task.

\begin{figure*}[t]
  \centering
  \begin{tabular}{c c c}
  \includegraphics[width=0.45\linewidth]{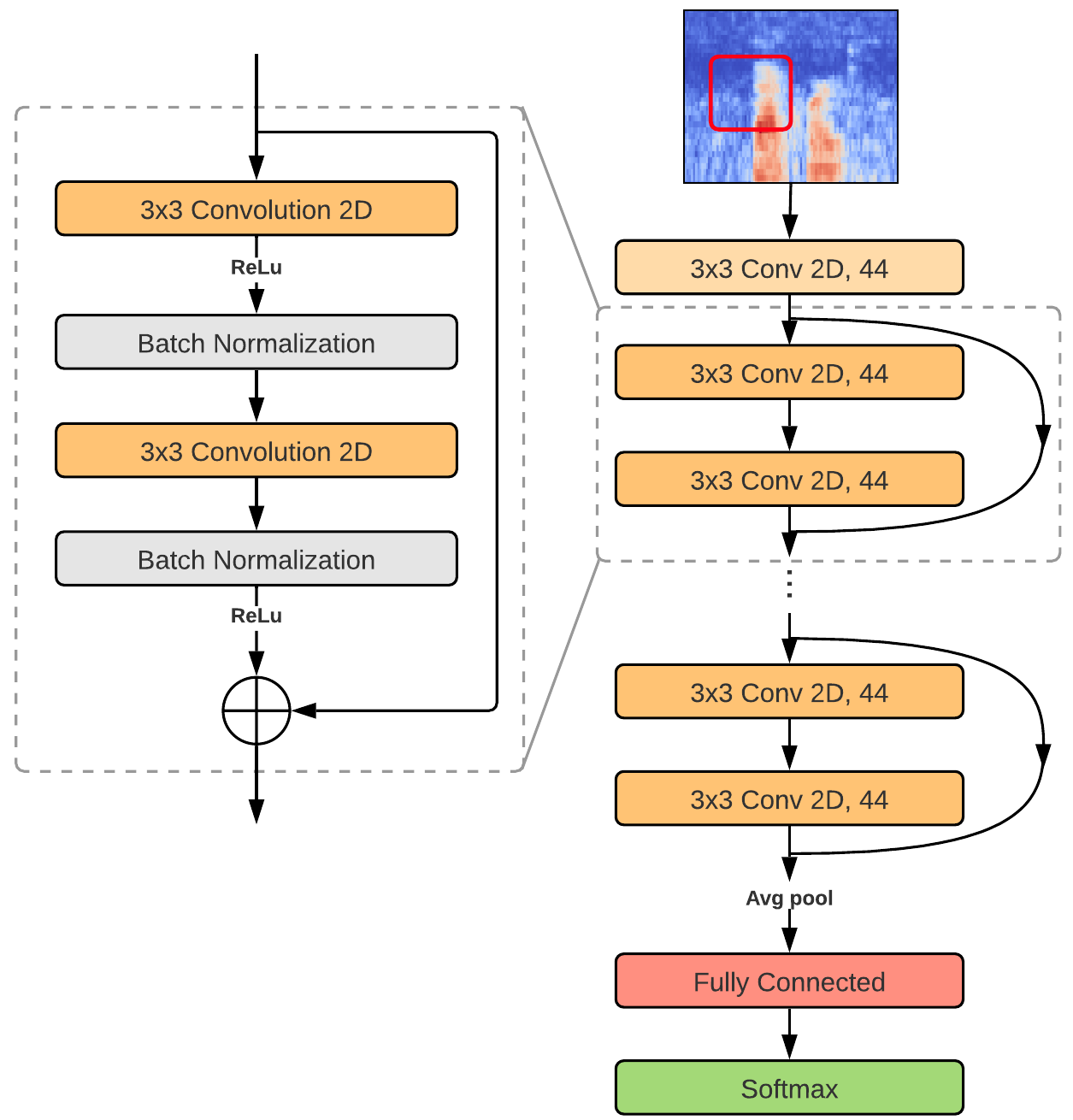} &
  \hfill &
  \includegraphics[width=0.45\linewidth]{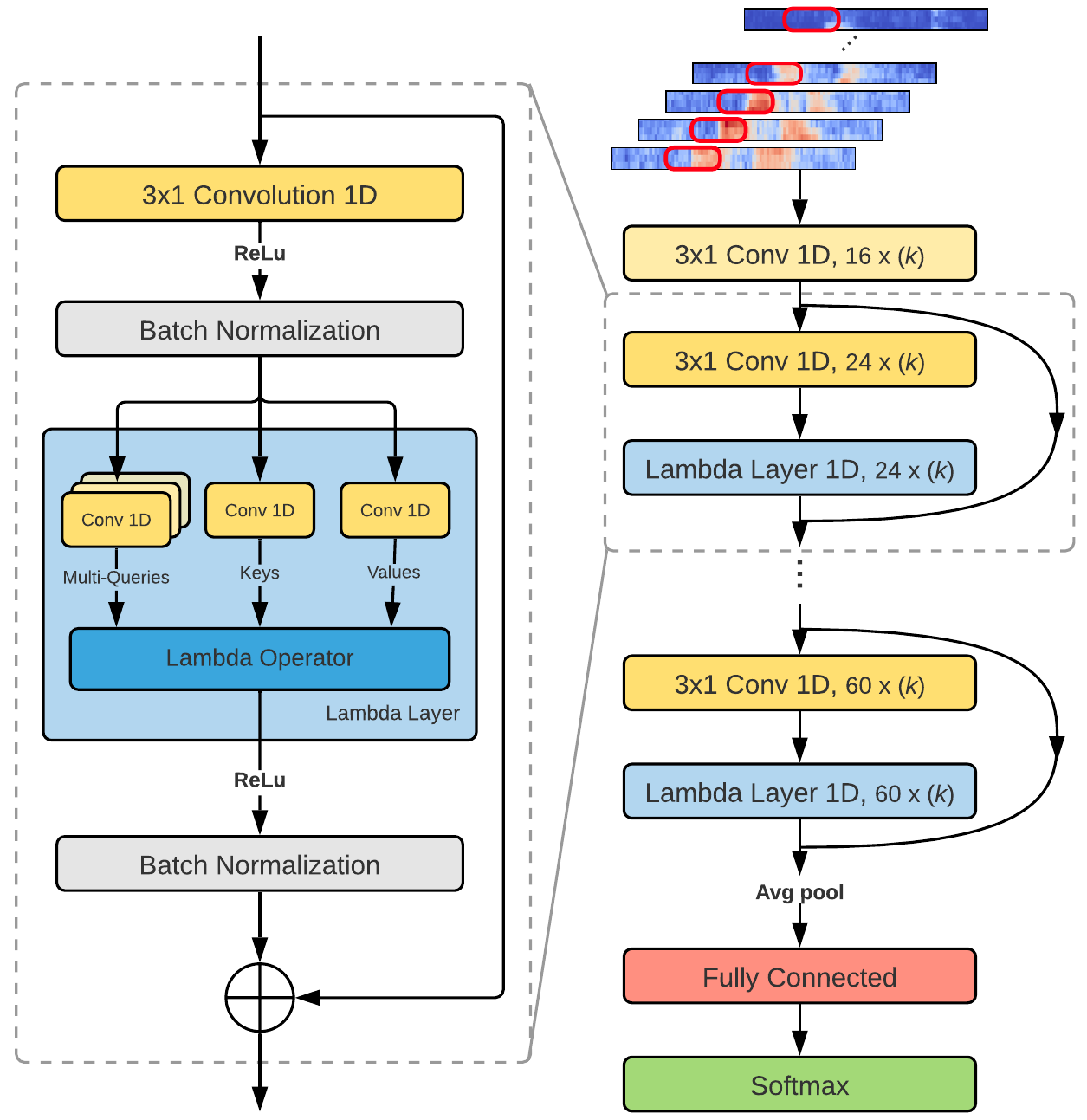} \\
  \footnotesize{\textit{ResNet15}} & & \footnotesize{\textit{LambdaResNet18}}
  \end{tabular}
  \caption{\small{Architectures compared for the Keyword Spotting task. Two-dimensional and uni-dimensional convolutional kernels are illustrated in the input of each model. \textbf{Left}: The \textit{ResNet15} architecture \cite{res15} is based on two-dimensional $3 \times 3$ convolutional filters with $44$ channels at each layer and no stride. Magnified residual block show the batch normalization and the usage of ReLu as activation function. \textbf{Right}: The \textit{LambdaResNet18} architecture built upon uni-dimensional $3 \times 1$ convolutional filters and Lambda layers in its residual blocks. Feature maps are expanded from $16$ to $60$ (with a multiplier $k$ for model variants) throughout the network with alternating stride values of $2$ and $1$. Residual Lambda block shows the operations inside the Lambda layer as well as the Batch Normalization and ReLu activation function.}}
  \label{fig:network}
\end{figure*}

\section{Proposed method}

\subsection{Input preprocessing}
Most of the CNN-based solutions for the Keyword Spotting task are based on pre-processing the audio into a two-dimensional input by transforming the speech signal to a Mel spectrogram representation. Although results using raw waveform without any Fourier pre-processing have also been investigated \cite{raw_kws, waveform_kws}, they do not show higher performance than using classical Mel spectrogram for the keyword spotting task. Other approaches have used the \textit{SincNet} architecture introduced by Ravanelli et al. \cite{sincnet} to pre-process the input audio, thus adding an additional front-end architecture \cite{sincnet_kws}.

Similar to the models we will compare our approach to, we use 40 mel filterbank energies to generate a classical log-scaled Mel spectrogram of the normalized raw audio input. Frames are estimated by using a $20$ms analysis window with a $10$ms stride. Note that instead of stacking all frequency bands, generating a two-dimensional input, that is, a classical spectrum image; we propose to understand each frequency band as uni-dimensional independent inputs, with as many channels as generated mels. A diagram of the proposed input can be seen in Figure \ref{fig:network}.

\subsection{Temporal Lambda layer}
Convolutional neural networks react to a small local area of the signal. The output feature map is generated from a linear combination of the inputs in the kernel window of the filter, learning local characteristics of the signal. The task of keyword spotting can benefit from learning long-range interactions but the attention map is very costly for the end task, usually carried in embedded devices with low computational and memory resources. The Lambda layer architecture, introduced by Bello \cite{lambda_networks}, aims to learn position-based interactions in global and local contexts to bring the benefits of the self-attention mechanisms in a computational lighter manner.


\subsubsection{Self-attention mechanism}

Presented by Vaswani et al. \cite{attention} under the Transformer architecture, self-attention mechanisms retrieves the relationship within the input sequence $X \in \mathbb{R}^{n \times d}$ by a set of Queries $Q \in \mathbb{R}^{n \times d_k} $, Keys $K \in \mathbb{R}^{n \times d_k}$ and Values $V \in \mathbb{R}^{n \times d_v}$ obtained through a linear projection of the input sequence. In the scaled dot-product attention, all queries and keys dot products are computed, scaled by $1/\sqrt{d_k}$ and normalized through the softmax operator $\sigma(\cdot)$ to obtain the attention map between all the input sequences. Then, this matrix weights the Values to obtain the final output $A \in \mathbb{R}^{n \times d_v}$, capturing long-range characteristics within the input sequence:
\begin{equation}
    A = \mathlarger{\sigma}\left(\frac{Q K^\top}{\sqrt{d_k}}\right)V
    \label{eq:attention}
\end{equation}


Instead of using a single attention function at each layer, it was found beneficial to use a multi-head attention architecture, where different projections of the input were learned in $h$ parallel layers to jointly attend information from different representations. The output of each attention layer (Equation \ref{eq:attention}) in the multi-head architecture, $A_1, \dots, A_h$, matrices are concatenated and linearly projected by $W_A \in \mathbb{R}^{h d_v \times d}$, generating the output $Y \in \mathbb{R}^{n \times d}$ of one layer of the Transformer: 

\begin{equation}
    Y = [ A_1 \,\;  A_2 \,\;\dots \,\; A_h ] W_A
    \label{eq:multihead}
\end{equation}

The multi-head attention is computationally expensive for two main reasons: the first one is the computation of a per-example attention map which is a \textit{quadratic} operation in terms of input size, see table~\ref{tab:complexity}, and secondly the time complexity scaling by a factor of $h$ in the multi-head architecture.

\subsubsection{Lambda operator}

The Lambda layer, presented and extensively analyzed by Bello \cite{lambda_networks}, avoids computing the expensive attention map by generating contextual lambda functions in the Lambda operator. Following the multi-head attention notation, $Q, K \in \mathbb{R}^{n \times d_k}$ are the Query and Key matrices and $V \in \mathbb{R}^{n \times d_v}$ is the Value matrix, each one obtained through a linear projection of the input. Then, the normalization step, through the softmax operator, is applied to the Key matrix across the context positions, $\sigma(K)$. The contextual lambda function for each input, $\lambda_n \in \mathbb{R}^{d_k \times d_v}$, to be applied to each query is defined as:

\begin{equation}
    \lambda_n = \underbrace{\sigma(K )^\top V}_{\lambda ^c} + \underbrace{E_{n}^\top V}_{\lambda^p_n}
\end{equation}

There are two terms in the contextual lambda expression: The content lambda, $\lambda^c$ encodes how to transform each query solely based on the context content. The position lambda $\lambda_n^p$ depends on the query position $n$ via the trainable positional embedding parameter $E_n \in \mathbb{R}^{n \times d_k}$. This latter lambda function encodes how to transform the query $q_n$ based on their relative positions.

Therefore, the output of the lambda layer, $Y \in \mathbb{R}^{n \times d_v}$, where each element $y_n$ of the output matrix is the application of each contextual lambda to its query, is expressed as:

\begin{equation}
    y_n = \lambda_n^\top q_n = (\lambda^c + \lambda_n^p)^\top q_n
    \label{eq:output_lambda}
\end{equation}

The columns of the lambda function $\lambda_n$ are interpreted as contextual aggregated features based on the content and position-based interactions. Applying the lambda function to the learned query distributes these contextual features and captures long-range interactions without the need of generating attention maps \cite{lambda_networks}.

Contrary to multi-head attention with $h$ parallel layers, where computation complexity scales by a factor of $h$, multi-query lambda reduces the computation by a factor by $h$ (being $h$ in this case the number of queries per lambda layer), leading to a better speed-accuracy trade-off than the Transformer architecture. Multi-query lambda follows the choice of $d_v = d / h$ and the generation of $h$ queries, $q_n^1, \dots, q_n^h$, such that the output $Y \in \mathbb{R}^{n \times d}$ is the concatenation of the different multi-query outputs:

\begin{equation}
    y_n = [ \lambda_n^\top q_n^1 \,\;   \lambda_n^\top q_n^2 \,\; \dots \,\; \lambda_n^\top q_n^h ] 
\end{equation}

\begin{table}[t]
  \caption{\small{Computational complexity comparison between multi-head self-attention and different multi-query Lambda variants. \textit{$b$: batch size, $h$: number of heads/queries, $n$: input length, $d$: output dimension, $d_k$: query/key dimension}, $r:$ local scope size.} \vspace{1mm}}
  \label{tab:complexity}
  \centering
  \begin{tabular}{lcc}
    \toprule
    \small{\textbf{\shortstack{Layer \\ Operation}}} & \textbf{\small \shortstack{Time \\ Complexity}} & \textbf{\small \shortstack{Space \\ Complexity}} \\
    \midrule
    \small{Self-Attention}    & \footnotesize{$\mathcal{O}(bn^2(h d_k + d))$}   & \footnotesize{$\mathcal{O}(bn^2h)$}  \\
    \small{Lambda Layer}      & \footnotesize{$\mathcal{O}(bn^2d_k (d/h))$}     & \footnotesize{$\mathcal{O}(n^2 d_k + b n d_k (d/h))$} \\
    \small{Lambda Conv.}      & \footnotesize{$\mathcal{O}(bnr d_k (d/h))$}     & \footnotesize{$\mathcal{O}(r d_k + b n d_k (d/h))$} \\
    \bottomrule
  \end{tabular}
\end{table}

Computing a single lambda output with reduced dimensionality, $\lambda_n \in \mathbb{R}^{dk \times d/h}$, and then concatenating the lambda-weighted queries in \textit{constant} time leads to a reduction of the time complexity by a factor of $h$.

The Lambda layer has still a \textit{quadratic} space and time complexity with respect to the input length due to the relative positional embedding. However, this term does not scale with the batch size (as in the attention operation). Computational and spatial complexity can be further improved when using Lambda convolutions. Despite the benefits of long-range interactions over the whole sequence, locality remains a strong inductive bias in the speech domain. Using global contexts may be excessive, as very long-time relations are usually not related between them. It may therefore be useful to restrict the scope of position interactions to a local area around the query position (similar to local self-attention). Therefore, rather than storing a position embedding for each value in the input sequence $n$, the positional embedding can be stored as a local scope of size $r$. As the computations are restricted to this scope, the lambda operator obtains \textit{linear} time complexity, resulting in a much more efficient computational alternative for capturing long-range local dependencies \cite{lambda_networks}. A comparison of time and spatial complexities is shown in Table \ref{tab:complexity}.



\subsection{Model Architecture}

The presented architecture is based on the \textit{ResNet15} model presented by Tang et al. \cite{res15}. This network is built upon residual connections \cite{ResNet} with two-dimensional convolutions filters. In the presented approach~\footnote{Code is available at: \url{https://github.com/Telefonica/LambdaNetwork}}, we reduce the dimensionality of the convolutional layers to one. By doing so, we do not only reduce the number of parameters in the networks, but the floating-point operations performed in inference are radically decreased. As the input of our model is a uni-dimensional mel band, uni-dimensional filters search for temporal patterns within that band. Similar architectures that replaced two-dimensional convolutional layers for uni-dimensional temporal convolutions had already been studied by Choi et al. \cite{ResNet1d} through the \textit{TC-ResNet14} architecture.


Solely reducing the dimensionality of the convolutional filters leads to a decrease in the model accuracy. To achieve similar metrics as two-dimensional architectures, networks based on uni-dimensional convolutions use wider kernels to capture longer temporal characteristics \cite{ResNet1d}, which increases the number of parameters of the model. To solve this concern, and following the idea of the Lambda Networks \cite{lambda_networks}, we replace the second uni-dimensional convolutional filter in the residual block with a temporal Lambda layer to capture these long dependencies. Therefore, our convolutional filters are $3\times1$ in size, smaller than other uni-dimensional architectures, as the long-term dependency is captured by the Temporal Lambda layer in the residual block.


The presented model, \textit{LambdaResNet18}, comprises eighteen layers: one uni-dimensional convolutional layer at the beginning, four residual layers with two residual Lambda blocks each one (with a convolutional and a Lambda layer each), and a fully connected layer followed by a Softmax activation. The proposed architecture is illustrated in Figure \ref{fig:network}. The number of channels follows a lighter implementation of the original \textit{ResNet} architecture and ranges from 16 in the first layer to 60 in the last one: $\{16, 24, 36, 48, 60\}$. Following the idea of \textit{TC-ResNet14-1.5}, we have implemented a variant of the model by increasing the number of channels by a factor of $k=2$, \textit{LambdaResNet18-2}, to study its scalability.

To reduce the spatial information while increasing the number of features map, a stride of $2$ is used at the first convolutional layer of each residual block. Following the \textit{ResNet} implementation, the shortcut connection of each block simply performs an identity mapping to match the dimensionality of the residual connection when needed.


The Lambda layer maps the input into queries, keys and values through a scalar non-linearity operator (self-attention input). Default implementation by Bello \cite{lambda_networks} was used in the temporal Lambda layer, that is a multi-query approach through $h=4$ heads, queries and keys dimension $d_k = 16$, value dimension $d_v = d/h$, being $d$ the input sequence size, and local context positional embedding of size $r = 23$. No intra-depth dimension was added in this architecture.
Finally, an average pooling is performed in the temporal dimension, resulting in an embedding of the input with frequency characteristics given by the long-range temporal relation captured by the Lambda layers. This embedding is then fed to the last fully connected classification layer, whose length varies depending on the output classes.

\section{Experiments}
\begin{table}[b]

  \caption{\small{List of the augmentation techniques used for speech perturbation (each one applied independently with a probability of \textit{p}).} \vspace{1mm}}
  \label{tab:augmentation}
  \centering
  \begin{tabular}{lcl}
    \toprule
    \textbf{Disturbance} & \textbf{\textit{p}} & \textbf{Parameters} \\
    \midrule
    Background noise    & $0.7$ & SNR range: $0$-$15$ dB  \\
    Clip distortion     & $0.2$ & Percentile threshold: $0.2$-$0.4$ \\
    Cropping            & $0.5$ & Cropping length: $10$-$100$ms \\
    Pitch Shift         & $0.3$ & Semitones range: $\pm4$ \\
    Temporal shift      & $0.3$ & Shift fraction: $\pm200$ms \\
    Temporal stretch    & $0.3$ & Stretching rate: $0.75$-$1.25$ \\
    Volume              & $0.5$ & Gain: $\pm5$ dB \\
    \bottomrule
  \end{tabular}
\end{table}

\begin{table*}[t]
  \caption{\small{Model comparison in terms of accuracy on the Google Speech Commands dataset for various data splits, number of parameters and number of multiplies measured in floating point operations (FLOPS). Accuracy metrics with $\star$ are extracted from the author's paper while the rest are trained using the same setup as our experiments. \vspace{2mm}}}
  \label{tab:accuracies}
  \centering
  \begin{tabular}{lccccc}
    \toprule
    \textbf{Models} & \textbf{35 [\%]} & \textbf{20 [\%]} & \textbf{10 [\%]} & \textbf{Parameters} & \textbf{Multiplies} \\
    \midrule
    Res15 \cite{res15}                  & $95.9$        & $96.3$        & $97.5$        & $237$K    & $894$ MFLOPS  \\
    Res15-Narrow \cite{res15}           & $-$           & $-$           & $94.0^\star$  & $42$K     & $160$ MFLOPS  \\
    Res15-Triplet \cite{res15_triplet}  & $96.4^\star$  & $-$           & $98.02^\star$ & $237$K    & $894$ MFLOPS  \\
    Att-RNN \cite{att_kws}              & $93.9^\star$  & $94.5^\star$  & $96.9^\star$  & $202$K    & $20.2$ MFLOPS \\
    MHAtt-RNN \cite{mhatt_rnn_kws}      & $-$           & $-$           & $98.0^\star$  & $743$K    & $21.2$ MFLOPS \\
    TC-Res14 \cite{ResNet1d}            & $91.3$        & $92.3$        & $95.4$        & $137$K    & $6.1$ MFLOPS \\
    TC-Res14-1.5 \cite{ResNet1d}        & $-$           & $-$           & $96.6^\star$  & $305$K    & $13.41$ MFLOPS \\
    
    KWT-1 \cite{kwt}        & $96.8^\star$           & $-$              & $97.7^\star$  & $607$K    & $-$ \\
    KWT-2 \cite{kwt}        & $97.5^\star$           & $-$              & $98.2^\star$  & $2394$K    & $-$ \\
    KWT-3 \cite{kwt}        & $97.5^\star$           & $-$              & $98.5^\star$  & $5361$K    & $-$ \\
    \midrule
    Lambda Res18                        & $93.1$        & $93.8$        & $96.7$        & $89$K     & $3.3$ MFLOPS \\
    Lambda Res18-2                      & $94.2$        & $94.5$        & $97.2$        & $270$K    & $8.4$ MFLOPS \\
    \bottomrule
  \end{tabular}
\end{table*}

\subsection{Experimental Setup}
Evaluations of the model were performed on the Google Speech Commands dataset \cite{google_speech_commands}. This dataset has been established as the common benchmark for the keyword spotting task as recent works use it for the training and evaluation of their models.
The Google Speech Commands dataset (in its newest version - v2) contains roughly $105$K one-second-long recordings of $35$ different keywords by thousands of different people. The dataset also provides different background noises to virtually augment the number of training samples at the pre-processing of the input training data.

It is common to split this dataset into different subtasks in order to widely evaluate the model in different scenarios. The difference in each one is the number of keywords detected. It is worth noting that not all the architectures we compare our approach to were reported in the subtasks presented.

\begin{itemize}[itemsep=1pt, topsep=1pt]
    \item \textbf{10 Keywords:} recognition among twelve classes: “yes”, “no”, “up”, “down”, “left”, “right”, “on”, “off”, “stop”, “go”, unknown or silence. 
    \item \textbf{20 Keywords:} recognition among the mentioned ten words in the previous subtask with an addition of the set of numbers ranging from “zero” to “nine”.
    \item \textbf{35 Keywords}: recognition on all the dataset classes.
\end{itemize}

Google Speech Commands dataset provides a SHA1-hashed list of audio speech filenames for the correct evaluation of the keyword spotting task. This split comprises 80\% of the data for training, 10\% for validation and 10\% for testing. 

We report the accuracies of the trained models for the different subtasks as well the number of parameters and floating-point operations performed. Receiver operating characteristic (ROC) curves, where the $x$-axis and $y$-axis show the false alarm rate and false reject rate, are plotted to compare the different models trained in the same environment. To extend the evaluating curve to capture multi-class metrics, we performed a micro-averaging over every class.

\subsection{Model Training}
The input to the model is a one-second-long keyword recording. We designed a data-augmentation module that applies a set of possible distortions to this input keyword waveform. This augmentation strategy is only present in the training process and the application of each perturbation is determined by random probability (see Table \ref{tab:augmentation}). Then, the resulting waveform is transformed to a log-scaled Mel spectrogram interpreted as a set of $40$ uni-dimensional vectors. The input shape is then $40$ different channels with a length of $100$ in the temporal axis. The model is trained through a Cross-Entropy loss function that updates the model parameters for every mini-batch of $256$ samples. Models are trained up to $200$ epochs and minimizing the loss over the validation data set. The reported metrics are computed over the test data. 

Stochastic gradient descend with a momentum of $0.9$ is used to train the network. The starting learning rate is $0.1$ and follows a cosine learning decay rate of $0.1$. We used a $\textit{L}_2$ weight decay of $10^{-3}$ as a regularization parameter. Background noises used in the speech augmentation are randomly sampled and cropped from the ones given by the Google Speech Commands dataset.

\begin{figure}[t]
  \centering
  \includegraphics[width=\linewidth]{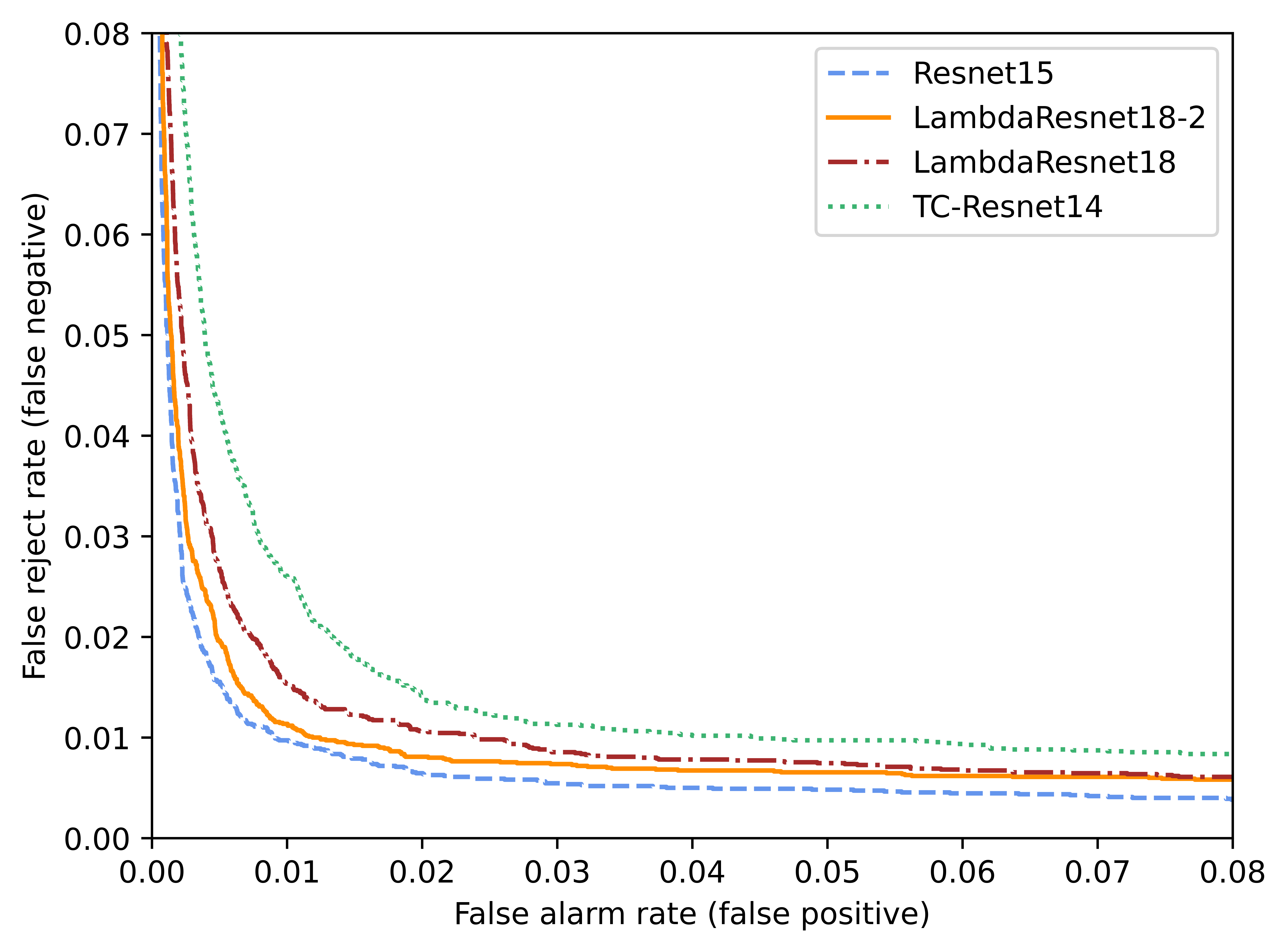}
  \caption{\small{Receiver operating characteristic (ROC) curves of the reported trained models in the same experimental setup.}}
  \label{fig:roc}

\end{figure}

\subsection{Results}

We trained four different models in the same environment: \textit{ResNet15}, \textit{TC-ResNet14} and the presented models \textit{LambdaResNet18} and \textit{LambdaResNet18-2}. Table \ref{tab:accuracies} shows the experimental results in terms of model accuracy for the three subtasks. Most of the architectures we compare our results to do not report all the subset metrics. We present all the possible results at each one of the dataset splits available for every model trained in our setup. Table \ref{tab:accuracies} also presents the number of parameters and computational complexity of the models. Both accuracy and model footprint metrics are equally important in keyword detection and they need to be considered together when evaluating a model.


In terms of accuracies, \textit{ResNet15} is the best model for the keyword spotting task in our experimental setup, but the architecture is $2.6\times$ larger than our solution and more than $100\times$ slower. On the other hand, if we compare our architecture to lighter models like \textit{TC-ResNet14}, we get a consistent $1.5\%$ of average improvement over all the subtasks, while being lighter and $2\times$ computationally faster. The presented \textit{LambdaResNet18} falls a bit behind the \textit{Att-RNN} model in terms of accuracy however our model is $56\%$ smaller in terms of parameters and $83\%$ computationally faster on the inference stage. The Keyword Transformer \textit{KWT} reports the best results in terms of accuracy but at the cost of being a much larger model, not being the optimal solution to integrate it in embedded devices with small computation capacity. Different version models varying the number of attention heads were proposed (\textit{KWT-1, KWT-2, KWT-3}). Our model is 85\% smaller than the lightest presented Keyword Transformer model, \textit{KWT-1} in which the number of multiplies was not reported. Table \ref{tab:complexity} shows that the complexity of a multi-query Lambda layer is significantly lower than the multi-headed self-attention approach used in the Keyword Transformer. Therefore, due to the large number of parameters in the KWS model, we can intuitively assume that the computational complexity of it is much higher than our solution.

We also trained a variant of our model that expands the channel dimension by a factor of $k=2$, \textit{LambdaResNet18-2}. This model increases the accuracy metrics at a cost of model parameters. Nevertheless, this architecture achieves very close state-of-the-art metrics while still being lighter and computationally faster than most of the presented models.

Receiver operating characteristic curves are plotted too for the 10 keyword subtask (Figure \ref{fig:roc}). The plotted curves are consistent with the results presented in Table \ref{tab:accuracies} where the \textit{ResNet15} architecture and the \textit{LambdaResNet18-2} get the lower curves  and \textit{LambdaResNet18} clearly surpasses previous uni-dimensional architectures like \textit{TC-ResNet14}.


\section{Conclusions and Future Work}

Through the keyword spotting task, we have shown the first steps of the Lambda framework for speech processing through the introduction of the \textit{LambdaResNet18} architecture. We proposed a light set of models in both footprint size and inference operations while keeping the accuracy on point with the state-of-the-art architectures. Moreover, the presented architecture can capture long-range interactions in a computationally efficient manner, thus being a perfect choice for their usage in embedded devices with very little computational resources.

Future research should consider the potential effects of training the presented models with new custom loss functions based on phonetic similarity which have shown to be beneficial for the keyword spotting task. In addition, the study of new variants of networks based on uni-dimensional convolution and attention models can also take this work as a baseline for future improvements on the keyword spotting task. Further applications of the Lambda architecture for more complex speech recognition problems will unleash new possibilities for lighter models that benefit from learning long-range interactions. 


\newpage
\bibliographystyle{IEEEbib}
\bibliography{refs}

\begin{thebibliography}{10}

\bibitem{speech_recognition_cnn}
Vineel Pratap, Qiantong Xu, Jacob Kahn, Gilad Avidov, Tatiana Likhomanenko,
  Awni Hannun, Vitaliy Liptchinsky, Gabriel Synnaeve, and Ronan Collobert,
\newblock ``Scaling up online speech recognition using convnets,''
\newblock {\em Annual Conference of the International Speech Communication
  Association (INTERSPECH)}, 2020.

\bibitem{attention}
Ashish Vaswani, Noam Shazeer, Niki Parmar, Jakob Uszkoreit, Llion Jones,
  Aidan~N. Gomez, Lukasz Kaiser, and Illia Polosukhin,
\newblock ``Attention is all you need,''
\newblock {\em Annual Conference on Neural Information Processing Systems
  (NIPS)}, 2017.

\bibitem{wav2vec2}
Alexei Baevski, Henry Zhou, Abdelrahman Mohamed, and Michael Auli,
\newblock ``wav2vec 2.0: A framework for self-supervised learning of speech
  representations,'' 2020.

\bibitem{speechBERT}
Yung-Sung Chuang, Chi-Liang Liu, Hung-Yi Lee, and Lin shan Lee,
\newblock ``Speechbert: An audio-and-text jointly learned language model for
  end-to-end spoken question answering,''
\newblock {\em Annual Conference of the International Speech Communication
  Association (INTERSPECH)}, 2020.

\bibitem{end2end_asr}
Gabriel Synnaeve, Qiantong Xu, Jacob Kahn, Tatiana Likhomanenko, Edouard Grave,
  Vineel Pratap, Anuroop Sriram, Vitaliy Liptchinsky, and Ronan Collobert,
\newblock ``End-to-end asr: from supervised to semi-supervised learning with
  modern architectures,''
\newblock {\em International Conference on Machine Learning (ICML)}, 2019.

\bibitem{transformers_asr}
Abdelrahman Mohamed, Dmytro Okhonko, and Luke Zettlemoyer,
\newblock ``Transformers with convolutional context for asr,'' 2019.

\bibitem{conformer}
Anmol Gulati, James Qin, Chung-Cheng Chiu, Niki Parmar, Yu~Zhang, Jiahui Yu,
  Wei Han, Shibo Wang, Zhengdong Zhang, Yonghui Wu, et~al.,
\newblock ``Conformer: Convolution-augmented transformer for speech
  recognition,''
\newblock {\em Annual Conference of the International Speech Communication
  Association (INTERSPECH)}, 2020.

\bibitem{dcase2021}
DCASE-2021,
\newblock ``Challenge on detection and classification of acoustic scenes and
  events (aasp),'' 2021.

\bibitem{cnn_kws}
T.~Sainath and C.~Parada,
\newblock ``Convolutional neural networks for small-footprint keyword
  spotting,''
\newblock {\em Annual Conference of the International Speech Communication
  Association (INTERSPECH)}, 2015.

\bibitem{ResNet1d}
Seungwoo Choi, Seokjun Seo, Beomjun Shin, Hyeongmin Byun, Martin Kersner,
  Beomsu Kim, Dongyoung Kim, and Sungjoo Ha,
\newblock ``Temporal convolution for real-time keyword spotting on mobile
  devices,''
\newblock {\em Annual Conference of the International Speech Communication
  Association (INTERSPECH)}, 2019.

\bibitem{aura}
David Bonet, Guillermo Cámbara, Fernando López, Pablo Gómez, Carlos Segura,
  and Jordi Luque,
\newblock ``Speech enhancement for wake-up-word detection in voice
  assistants,'' 2021.

\bibitem{lambda_networks}
Irwan Bello,
\newblock ``Lambda networks: Modeling long-range interactions without
  attention,''
\newblock {\em International Conference in Learning Representations (ICLR)},
  2021.

\bibitem{kwt}
Axel Berg, Mark O'Connor, and Miguel~Tairum Cruz,
\newblock ``Keyword transformer: A self-attention model for keyword spotting,''
  2021.

\bibitem{multiband}
C.~Cerisara, J.-P. Haton, J.-F. Mari, and D.~Fohr,
\newblock ``A recombination model for multi-band speech recognition,''
\newblock in {\em IEEE International Conference on Acoustics, Speech and Signal
  Processing (ICASSP)}, 1998, vol.~2.

\bibitem{temporal_conv_kws}
Ximin Li, Xiaodong Wei, and Xiaowei Qin,
\newblock ``Small-footprint keyword spotting with multi-scale temporal
  convolution,'' 2020.

\bibitem{res15}
R.~{Tang} and J.~{Lin},
\newblock ``Deep residual learning for small-footprint keyword spotting,''
\newblock in {\em International Conference on Acoustics, Speech and Signal
  Processing (ICASSP)}, 2018, pp. 5484--5488.

\bibitem{mlp_kws}
G.~{Chen}, C.~{Parada}, and G.~{Heigold},
\newblock ``Small-footprint keyword spotting using deep neural networks,''
\newblock in {\em International Conference on Acoustics, Speech and Signal
  Processing (ICASSP)}, 2014, pp. 4087--4091.

\bibitem{mlp_kws_temp}
Zhiming Wang, Xiaolong Li, and Jun Zhou,
\newblock ``Small-footprint keyword spotting using deep neural network and
  connectionist temporal classifier,'' 2017.

\bibitem{hmm}
J.~R. {Rohlicek}, W.~{Russell}, S.~{Roukos}, and H.~{Gish},
\newblock ``Continuous hidden markov modeling for speaker-independent word
  spotting,''
\newblock in {\em International Conference on Acoustics, Speech and Signal
  Processing (ICASSP)}, 1989, pp. 627--630.

\bibitem{rnn_kws}
Sercan~O. Arik, Markus Kliegl, Rewon Child, Joel Hestness, Andrew Gibiansky,
  Chris Fougner, Ryan Prenger, and Adam Coates,
\newblock ``Convolutional recurrent neural networks for small-footprint keyword
  spotting,''
\newblock {\em Annual Conference of the International Speech Communication
  Association (INTERSPECH)}, 2017.

\bibitem{att_kws}
Douglas~Coimbra de~Andrade, Sabato Leo, Martin Loesener Da~Silva Viana, and
  Christoph Bernkopf,
\newblock ``A neural attention model for speech command recognition,'' 2018.

\bibitem{mhatt_rnn_kws}
Oleg Rybakov, Natasha Kononenko, Niranjan Subrahmanya, Mirkó Visontai, and
  Stella Laurenzo,
\newblock ``Streaming keyword spotting on mobile devices,''
\newblock {\em Annual Conference of the International Speech Communication
  Association (INTERSPECH)}, 2020.

\bibitem{res15_triplet}
Roman Vygon and Nikolay Mikhaylovskiy,
\newblock ``Learning efficient representations for keyword spotting with
  triplet loss,'' 2021.

\bibitem{ViT}
Alexey Dosovitskiy, Lucas Beyer, Alexander Kolesnikov, Dirk Weissenborn,
  Xiaohua Zhai, Thomas Unterthiner, Mostafa Dehghani, Matthias Minderer, Georg
  Heigold, Sylvain Gelly, Jakob Uszkoreit, and Neil Houlsby,
\newblock ``An image is worth 16x16 words: Transformers for image recognition
  at scale,''
\newblock in {\em International Conference on Learning Representations (ICLR)},
  2021.

\bibitem{raw_kws}
K.~{Kumatani}, S.~{Panchapagesan}, M.~{Wu}, M.~{Kim}, N.~{Strom}, G.~{Tiwari},
  and A.~{Mandai},
\newblock ``Direct modeling of raw audio with dnns for wake word detection,''
\newblock in {\em Automatic Speech Recognition and Understanding Workshop
  (ASRU)}, 2017, pp. 252--257.

\bibitem{waveform_kws}
Patrick Jansson,
\newblock ``Single-word speech recognition with convolutional neural networks
  on raw waveforms,''
\newblock M.S. thesis, Arcada University of applied sciences, 2018.

\bibitem{sincnet}
Mirco Ravanelli and Y.~Bengio,
\newblock ``Speaker recognition from raw waveform with sincnet,''
\newblock {\em Spoken Language Technology Workshop (SLT)}, 2018.

\bibitem{sincnet_kws}
Simon Mittermaier, Ludwig Kürzinger, Bernd Waschneck, and Gerhard Rigoll,
\newblock ``Small-footprint keyword spotting on raw audio data with
  sinc-convolutions,''
\newblock in {\em International Conference on Acoustics, Speech and Signal
  Processing (ICASSP)}, 2020, pp. 7454--7458.

\bibitem{ResNet}
K.~{He}, X.~{Zhang}, S.~{Ren}, and J.~{Sun},
\newblock ``Deep residual learning for image recognition,''
\newblock in {\em Conference on Computer Vision and Pattern Recognition
  (CVPR)}, 2016, pp. 770--778.

\bibitem{google_speech_commands}
Pete Warden,
\newblock ``Speech commands: A dataset for limited-vocabulary speech
  recognition,'' 2018.

\end{thebibliography}

\end{document}